\documentclass[aps,prl,twocolumn,showpacs,10pt]{revtex4-1}
\usepackage{graphicx}
\usepackage{amssymb}
\usepackage{amsmath}
\usepackage{comment}
\usepackage{footmisc}
\usepackage{color}
\pdfoutput=1

\usepackage{ulem}
\newcommand{\away}[1]
{}

\begin{document}

\title{
Topological  Floquet Phases  in Driven Coupled Rashba Nanowires\\
}

\author{Jelena Klinovaja$^1$, Peter Stano$^2$, and Daniel Loss$^{1,2}$}
\affiliation{$^1$Department of Physics, University of Basel, Klingelbergstrasse 82, CH-4056 Basel, Switzerland\\
$^2$RIKEN Center for Emergent Matter Science, Wako, Saitama 351-0198, Japan }

\begin{abstract}
We consider   periodically-driven  arrays of weakly coupled wires with  conduction and valence  bands of Rashba type and study the resulting Floquet states. This non-equilibrium system can be tuned into non-trivial phases such as of  topological insulators, Weyl semimetals, and dispersionless zero-energy edge mode regimes. In the presence of strong electron-electron interactions, we generalize these regimes to the fractional case, where elementary excitations have fractional charges $e/m$ with $m$ being an odd integer.
\end{abstract}



\maketitle

\textit{Introduction.} 
Topological effects in condensed matter systems have been at the center of attention for many years. From quantum Hall effect over topological insulators (TIs) \cite{bib:Kane2010,bib:Zhang2011,bib:Hankiewicz2013,bib:Pankratov1985,
bib:Volkov1987,bib:Zhang2006,bib:Zhang2007,bib:Zhang2009,Nagaosa_2009,bib:Moler2012,Amir_ext,Leo_exp} and Weyl semimetals \cite{W_1,W_2,W_3,W_4,W_Sasha}, 
to exotic bound states such as Majorana fermions \cite{MF_Sato,MF_Sarma,MF_Oreg,alicea_majoranas_2010,MF_ee_Suhas,
potter_majoranas_2011,Klinovaja_CNT,Pascal,bilayer_MF_2012,Bena_MF,Rotating_field,
Ali,RKKY_Basel,RKKY_Simon,RKKY_Franz,mourik_signatures_2012,deng_observation_2012,
das_evidence_2012,Rokhinson,Goldhaber,marcus_MF,Ali_exp,Basel_exp} and parafemions~\cite{PF_Linder,PF_Clarke,PF_Cheng,Ady_FMF,PF_Mong,vaezi_2,PFs_Loss,
PFs_Loss_2,PFs_TI,barkeshli_2},
the interest is driven  both by fundamental physics and the promise for topological quantum computation. Despite the fact that there are many proposals by now for observing such topological effects in experiments the search for the most optimal system 
still continues unabated.  

While most of these studies  were
 focused on static structures, it has recently been proposed to extend the concept of topological phases to non-equilibrium systems, described by Floquet states \cite{Fl_Oka,Fl_Demler,Fl_Nature_Linder,Fl_PRB_Linder,Tanaka_PRL,Fl_Rudner,Fl_Liu,Platero_2013,Fl_Reynoso,Fl_Grushin,Sen_MF}. 
Remarkably, this approach no longer relies on given material properties, such as strong spin orbit interactions (SOI), typically necessary for reaching topological regimes, but
instead  allows one to turn initially non-topological materials such as graphene \cite{Fl_Oka} and non-band-inverted semiconducting wells into TIs \cite{Fl_Nature_Linder} by applying an external driving field. 

An even bigger challenge is to describe topological effects that involve fractional excitations. This requires the presence of strong electron-electron interactions. 
However,  given the difficulties in the search of conventional TIs, it would be even more surprising to expect such phases to occur naturally.
Moreover, even if they existed, two-dimensional (2D) systems with electron-electron interactions are difficult to describe analytically and often progress can come only from numerics \cite{Fl_Grushin}. 

\begin{figure}
\includegraphics[width=0.85\columnwidth]{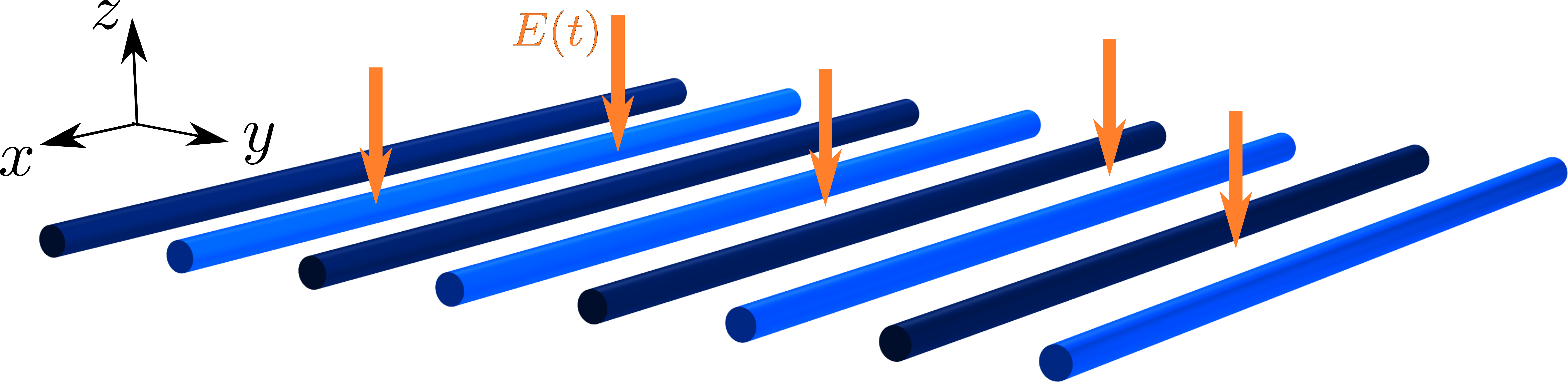}
\caption{A periodic array of weakly coupled Rashba wires (blue cylinders) aligned in the $x$ direction in the $xy$ plane. The sign of  SOI changes from positive  (dark blue) to negative (light blue) inside the unit cell composed of two (four) wires in the first (second) model, see below. The driving field $E(t)$ with period $\omega$ applied along $z$ (orange arrows) results in a coupling of the conduction and valence bands (separated by a gap $\Delta_g$)
at resonance when $\hbar \omega=\Delta_g$.}
\label{model}
\end{figure}

Here, we circumvent this difficulty by considering strongly anisotropic 2D systems \cite{Lebed,Yakovenko_PRB,Kane_PRL,Stripes_PRL} formed by weakly coupled Rashba wires (see Fig. \ref{model}), where each of them can  be  treated as a one-dimensional Luttinger liquid by bosonization 
\cite{Kane_PRB,Stripes_arxiv,Stripes_nuclear,yaroslav,Neupert,oreg_2,Oreg,tobias_1,AQHE,tobias_2,sela,Gutman}.  
This will allow us to introduce the Floquet version not only of TIs but also of Weyl semimetals in driven 2D systems. Importantly, in this way we can also address fractional regimes and are able to obtain the Floquet version of  fractional TIs and Weyl semimetals.

\begin{figure}[!t]
\includegraphics[width=\columnwidth]{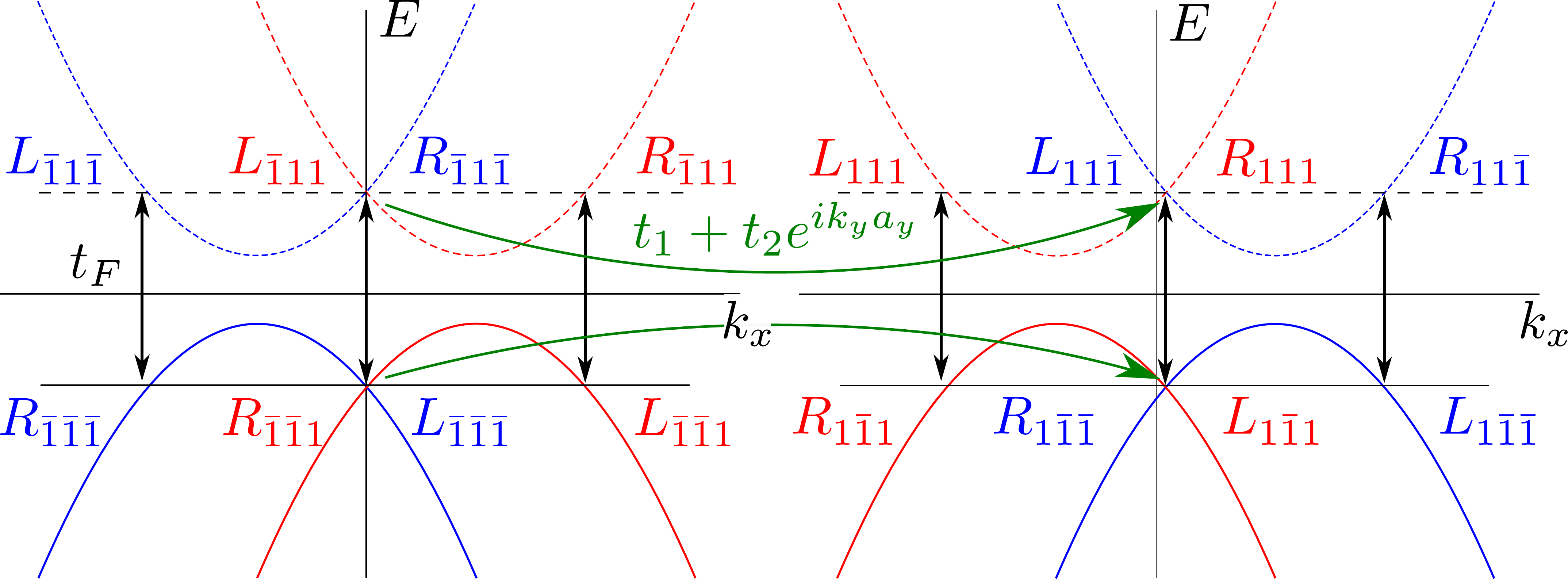}
\caption{The unit cell of the 2D system composed of two Rashba wires  with opposite SOI is shown in $k_y$-momentum  space,
where $R_{\lambda\eta\sigma}\equiv R_{k_y\lambda\eta\sigma}$ {\it etc}.   The  gap $\Delta_g$ separates the valence ($\eta=\bar  1$)  and conduction ($\eta=1$) bands. 
 The index $\sigma=1$ ($\sigma=\bar 1$) refers to the spin up (spin down) band shown in red (blue).  The chemical potential $\mu$  is tuned to  $E_{so}$
 and the driving frequency is chosen as $\hbar \omega =\Delta_g$, resulting in  resonant scattering between  bands with amplitude $t_F$ (vertical black arrows), opening gaps. The inter-wire tunneling with amplitude $t_1+t_2 e^{ik_y a_y}$ is shown by green arrows. }
\label{fig:2D_model}
\end{figure}

\textit{Topological dispersionless edge modes in driven systems.} 
We consider a 2D model formed by weakly tunnel-coupled Rashba wires, see 
Fig. \ref{model}. The spectrum of such one-dimensional wires consists of 
conduction 
and valence
 bands with Rashba SOI, separated by a gap $\Delta_g$, and labeled by the index $\eta=\pm 1$, 
  see Fig. \ref{fig:2D_model}.  Semiconducting wires, atomic chains, as well as graphene and metal dichalcogenide nanoribbons can be used. However, for simplicity, we refer to all such one-dimensional or single channel systems as wires in this work. The wires are aligned along the $x$ axes and lie in the $xy$ plane. The unit cells labeled by the index $n$ are composed of two wires with opposite SOI labeled by the index $\lambda=\pm 1$.  
The  kinetic part of the  Hamiltonian (density) corresponding to the $\lambda$ wire in the $n$th unit cell is written as $H_{0n\lambda} =\sum_{\eta\sigma} \eta \Psi_{n\lambda \eta \sigma }^\dagger   \left(\delta_{1\eta}\Delta_g-\frac{\hbar^2 \partial_x^2}{2 m_0 } + \alpha \lambda \sigma \partial_x  \right)\Psi_{n \lambda\eta \sigma }$.
Here, $m_0$ is the effective mass, and $\alpha$ is the strength of the SOI, with corresponding wavevector $k_{so}=m_0 \alpha/\hbar^2$ and energy $E_{so}=\hbar^2 k_{so}^2/2m_0$. Without loss of generality, we choose the SOI vector to point in $z$ direction. Here, $\Psi_{n\lambda\eta \sigma }(x)$  is the annihilation operator acting on the particle of the $\eta$ band with spin $\sigma$, and located at position $x$ of  the $\lambda$ wire in the $n$th unit cell. 
We assume the chemical potential $\mu$ to be tuned to the SOI induced crossing of the valence bands.

Next, we allow for an oscillating electric field, at frequency $\omega$, either due to an external electromagnetic radiation, or induced by periodically driving a voltage difference between a back- and top gate  enclosing the wire array.
Driving at resonance across the band gap, that is with $\hbar \omega = \Delta_g$, opens a dynamical gap \cite{faisal1997:PRA,wang2013:S}, essential for inducing non-trivial topology  \cite{Fl_Nature_Linder,kitagawa2011:PRB,Fl_Rudner}. Using the Floquet representation \cite{review,sambe1973:PRA,PACTs}, this gap arises as a splitting of degeneracies in the quasi-energy spectrum of the Floquet operator. It corresponds to the lowest order of the degenerate perturbation theory in the electric field amplitude, or, in another words, to a single photon emission/absorption process.  In this approximation, the dipole matrix element plays then the role of an effective tunneling amplitude which connects the two bands in resonance as described next.

To simplify calculations, we work in the regime of strong SOI and linearize the problem \cite{Bernd_1,Composite_MF} by representing operators in terms of spatially slowly-varying left and right mover fields defined around the Fermi points $k_F=0,\pm 2 k_{so}$  (see Fig.~\ref{fig:2D_model}) as
$\Psi_{n\lambda \eta \sigma} = R_{n\lambda \eta \sigma}(x) e^{i  k_{so} (\eta -\sigma \lambda)x} + L _{n\lambda \eta \sigma } (x) e^{-i   k_{so} (\eta + \sigma \lambda)x}$.
The system is translation invariant in  $y$ direction, so we introduce the conserved momentum $k_y$ via $\Psi_{n} = \sum_{k_y} e^{i n k_y a_y} \Psi_{k_y} $, where $a_y$ is the unit cell size. The kinetic term becomes $H_{0} =i \hbar \upsilon_F \sum_{k_y \lambda \eta \sigma } (L^\dagger_{k_y\lambda \eta \sigma } \partial_x L_{k_y\lambda \eta \sigma }- R^\dagger_{k_y\lambda \eta \sigma } \partial_x R_{k_y\lambda \eta \sigma })$
with $\upsilon_F$ being the Fermi velocity.
The inter-wire tunneling term, 
 $H_{t} = \sum_{n \eta \sigma} (t_1 \Psi_{n 1 \eta \sigma}^\dagger  \Psi_{n \bar 1  \eta \sigma} + t_2 \Psi_{(n+1) 1 \eta \sigma}^\dagger  \Psi_{n \bar 1  \eta \sigma}) +{\rm H.c.}$,
 becomes in linearized form
$H_{t}= \sum_{k_y\sigma} t_y  e^{i\phi} \big( L^\dagger_{k_y 1 \bar \sigma  { \sigma}} R_{k_y  \bar 1 \bar \sigma   \sigma} +R^\dagger_{ k_y  1\sigma \sigma} L_{k_y  \bar 1 \sigma \sigma} ) + {\rm H.c.}$,
where we keep only slowly oscillating terms.
Here, we introduced $(t_1+t_2 e^{ik_y a}) \equiv t_y e^{i\phi}$ with 
$t_y=\sqrt{t_1^2+t_2^2+2t_1t_2\cos (k_y a_y)}$.
The magnetic field is  applied in-plane and perpendicular to the SOI vector, say, in  $x$ direction and is described by 
$H_Z=\Delta_Z \sum_{\eta \sigma \sigma' n\lambda} \Psi_{n\lambda\eta \sigma }^\dagger  (\sigma_x)_{\sigma\sigma'}\Psi_{n\lambda \eta \sigma' }$, yielding the
 linearized version 
\begin{align}
&H_Z= \Delta_Z \sum_{k_y \lambda \eta }  R^\dagger_{ k_y  \lambda \eta (\lambda \cdot\eta)} L_{k_y  \lambda \eta {{(\lambda\cdot {\bar\eta})}}}  +{\rm H.c.}
\end{align}

The time-dependent driving term couples two bands  and is given by $H_{F}=  t_F \sum_{n \lambda \sigma} \Psi_{n\lambda 1 \sigma}^\dagger  \Psi_{n \lambda \bar 1 \sigma} +{\rm H.c.} $, or after linearization by
\begin{align}
&H_{F}= 
 t_F \sum_{n \lambda \eta \sigma } R^\dagger_{ n \lambda \eta \sigma} L_{n \lambda\bar \eta \sigma  } 
 + {\rm H.c.},
\end{align}
where we neglected the fast-oscillating terms and the weak photon-assisted inter-wire inter-subband scattering. The amplitude $t_F$ 
 is proportional to the inter-band dipole matrix element and the $E$-field amplitude, see Ref.  \cite{SM}. The weak driving which we consider corresponds to $t_F \ll \Delta_g$.

\begin{figure}[!b]
\includegraphics[width=0.8\columnwidth]{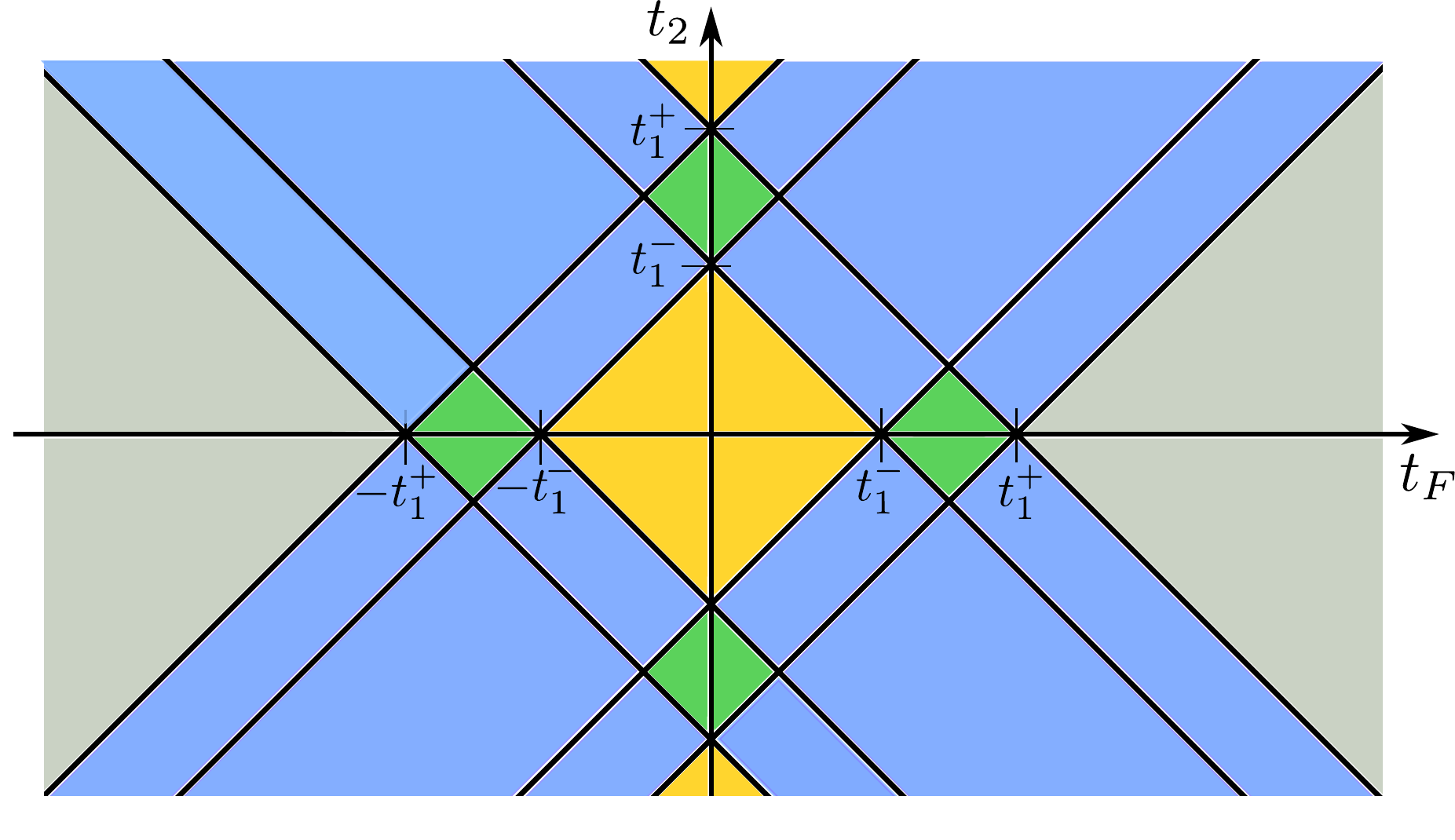}
\caption{The phase diagram of 
the model shown in Fig.  \ref{fig:2D_model} as a function of the Floquet coupling $t_F$ and the tunneling amplitude $t_2$.   
For fixed values of  the tunneling amplitude  $t_1$ and the Zeeman energy $\Delta_Z$ we determine $t_1^\pm = t_1\pm \Delta_Z$ (here, we assume that $\Delta_Z<t_1$ such that $t_1^->0$). The system in the topological phase is gapped in the bulk and hosts either one zero-energy edge mode (green area) or two zero-energy edge modes (yellow area). In contrast, in the trivial phase, the system does not support edge modes in the gap (gray area). In the Weyl phase (blue area), edge modes connect two gapless bulk cones, see Fig.~\ref{fig:phase_2}b.}
\label{fig:phase_diagramm}
\end{figure}

\begin{figure}[!t]
\includegraphics[width=0.85\columnwidth]{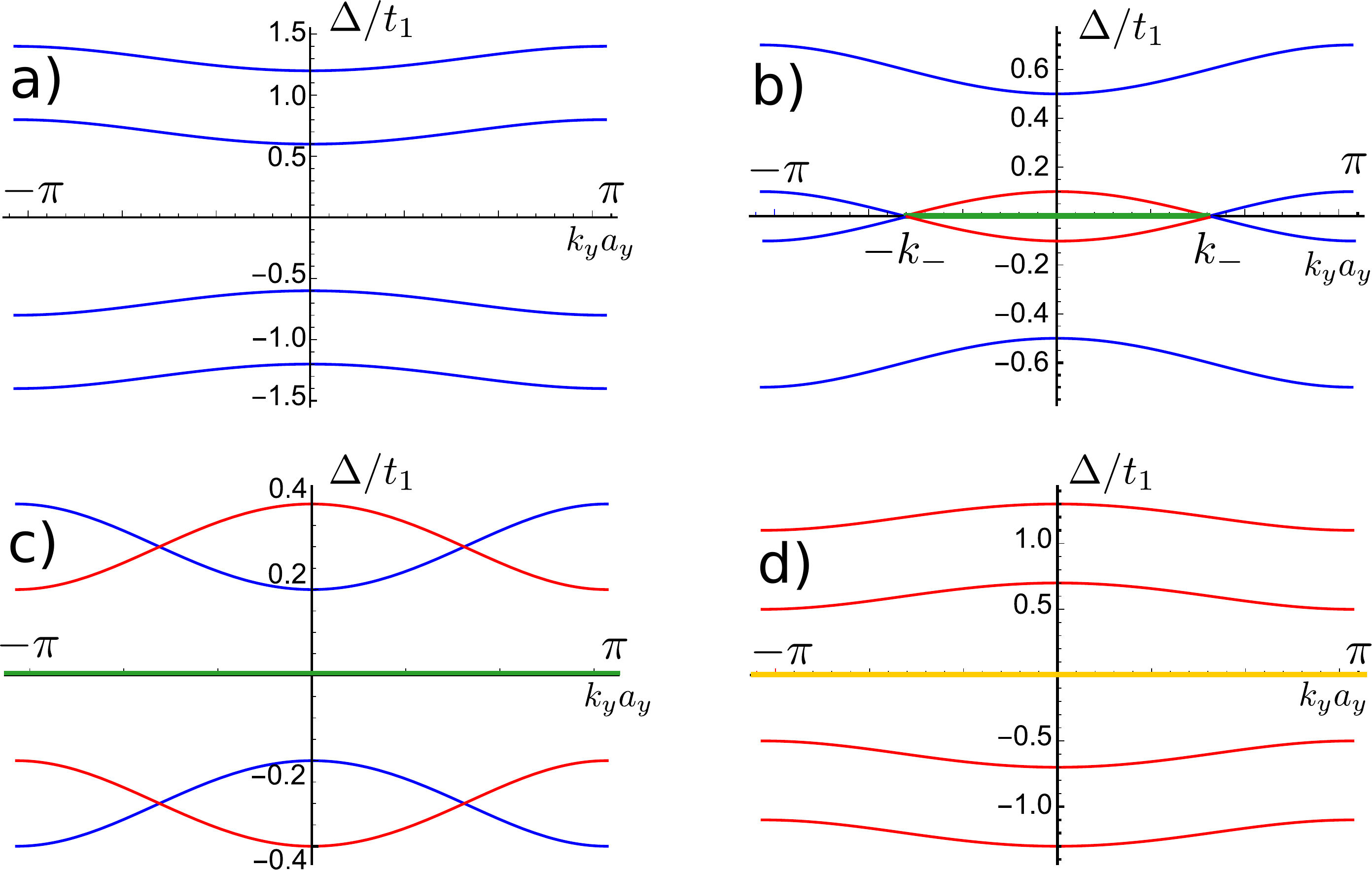}
\caption{The dependence of the minimum bulk gap $\Delta$ of $E_{2\pm\pm}$ occurring at $k_x=0$ on the transverse momentum $k_y$. The parameters are chosen as $t_2/t_1=0.1$, $\Delta_Z/t_1=0.3$. (a) The system is in the trivial phase for $t_F/t_1=2$.  (b) As  $t_F$ decreases, $t_F/t_1=1.4$, the bulk gap closes at $k_y=\pm k_{-}$, and a single zero-energy mode at each edge (green line) connects two Dirac cones (bulk spectrum). (c) For  $t_F/t_1=1$, the bulk gap  reopens and leaves one of the bands (red lines) in the topological phase with a single zero-energy mode at each edge. (d) At small $t_F$, $t_F/t_1=0.1$, both bands are in the topological phase resulting in two zero-energy modes at each edge (yellow line). }
\label{fig:phase_2}
\end{figure}

The bulk spectrum is given by $E_{1}^2= (\hbar \upsilon_F k_x)^2 +t_F^2$ and $E_{2\pm\pm}^2=(\hbar \upsilon_F k_x)^2 +\left(t_F\pm t_y \pm \Delta_Z\right )^2$.
The system is gapless in the bulk if $t_y=\pm (t_F\pm \Delta_Z)$. For simplicity, we assume henceforth that all amplitudes are non-negative.
There are two pairs of dispersionless  zero-energy modes for $t_F<|t_y-\Delta_Z|$. If $|t_y-\Delta_Z|<t_F<|t_y+\Delta_Z|$, there is only one zero-energy mode. The system is trivial if $t_F>|t_y+\Delta_Z|$. 
The phase diagram recalculated in terms of the tunneling amplitudes $t_{1}$ and $t_{2}$ is shown in Fig.~\ref{fig:phase_diagramm}. Interestingly, there emerges also a topological phase similar to Weyl semimetals \cite{W_1,W_2,W_3,W_4,W_Sasha}, see Fig.~\ref{fig:phase_2}. Indeed, the spectrum is gapless due to two (possibly four) Dirac cones located at $\pm k_{\pm}$, defined by the condition for closing  the bulk gap, $\cos (k_{\pm} a) = [(t_F\pm \Delta_Z)^2-t_1^2-t_2^2]/2t_1t_2$. 
In addition, there are dispersionless zero-energy edge modes (Fermi arcs) that connect the Dirac cones, see e.g. Fig.~\ref{fig:phase_2}b.
For details on the wavefunctions we refer to the Supplemental Material \cite{SM}, where we also discuss the generalization of the model to the {\it fractional} Weyl semimetal regime.

\begin{figure*}[!t]
\includegraphics[width=1.7\columnwidth]{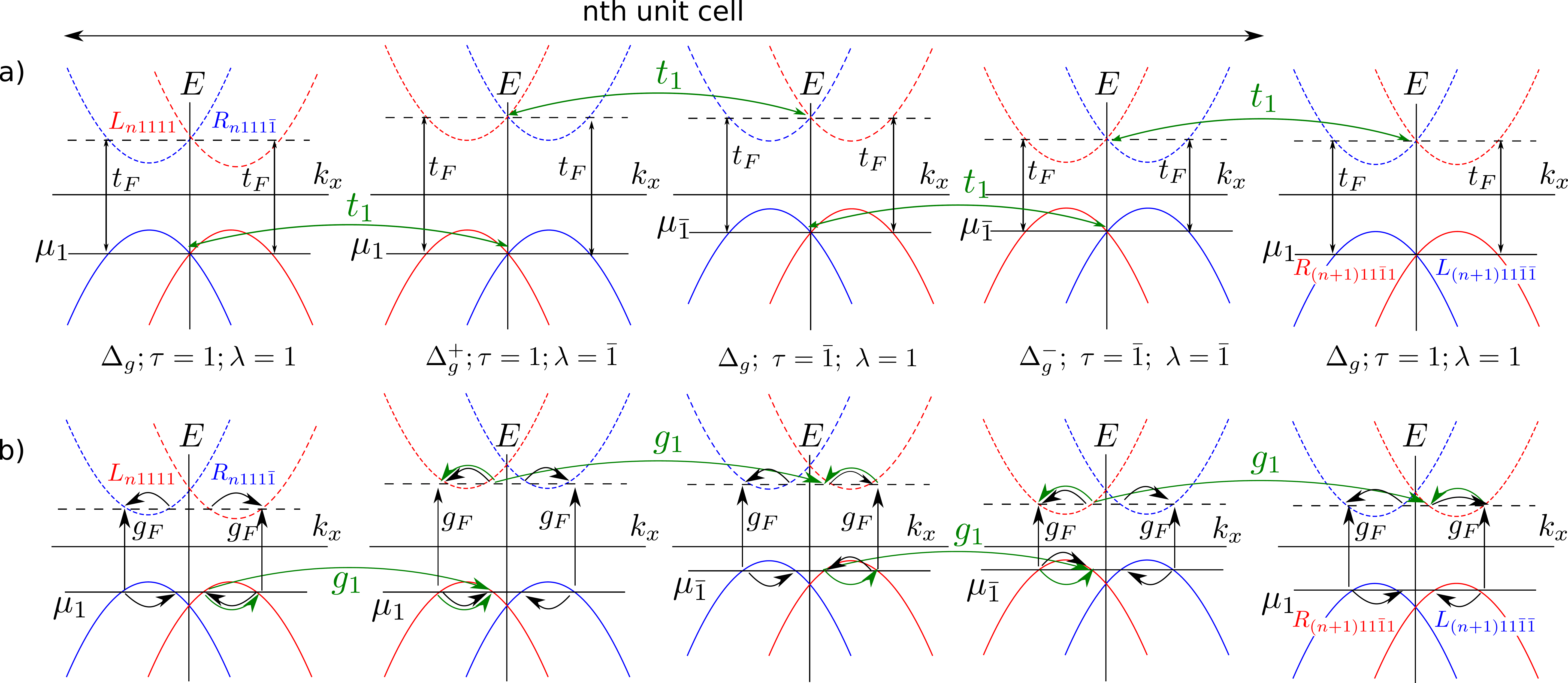}
\caption{The unit cell of a Floquet TI, consisting of four wires labeled by $\tau, \lambda=\pm 1$. The sign of the SOI  alternates between wires while the band gap alternates between $\Delta_g$ and $\Delta_{g\pm} = \Delta_g \pm \delta$. (a) In the integer regime, 
$\mu_\tau$ is tuned to $E_{so}$ and the driving frequencies are chosen to match the band gaps. The interwire hopping of  amplitude $t_1$ opens gaps at $k_x=0$ if resonant. The remaining gaps of size $t_F$ are opened by the time-dependent driving. 
At each edge of the array, there is a pair of helical modes left gapless. For example, if the system is terminated as shown in this figure, the modes $L_{n1111}$ and $R_{n111\bar 1}$ at the left edge stay gapless. (b) In the fractional regime,  $\mu_{\tau}$ is shifted to $E_{so\lambda}/9$ and the driving frequency is re-adjusted accordingly.
The leading term in the tunneling between wires $H_1^{ee}$ (green arrows) of  amplitude $g_1$ involves two back-scattering events and opens a partial  gap  due to strong  interactions. The driving term $H_F^{ee}$ of  amplitude $g_F$ (black arrows) commutes with $H_1^{ee}$ and both can order simultaneously.}
\label{2D}
\end{figure*}

{\it Floquet TI.} The model considered above hosts only zero-energy edge modes. In our second model we propose a setup that will host helical edge modes. To achieve this, we need to break the symmetry between hoppings along the edge in positive and negative directions. This can be implemented in a model based on a unit cell consisting of four wires, labeled by two indices $\lambda,\tau=\pm 1$, see Fig.~\ref{2D}. The sign of the Rashba SOI $\alpha_{\lambda}\equiv \lambda \alpha$ alternates between wires as well as the band gap 
$\Delta_{g\tau\lambda} =\Delta_{g} + \tau  (1-\lambda)\delta /2$. The chemical potential is tuned to the SOI energy  inside each wire and the driving frequency 
matches the resonance between the Fermi level and  the conduction band with the same momentum,  $\hbar \omega_{\tau\lambda} = \Delta_{g\tau\lambda} + 2 E_{so}$, see Fig.~\ref{2D}. We note that the bias between the wires due to different chemical potentials results in a small leakage current, which we estimate in Ref. \cite{SM}. Due to translation invariance along the wire, the inter-wire tunneling is assumed to be momentum-and spin-conserving. 
For simplicity, we assume that all inter-wire tunneling amplitudes are equal. In the limit $t_1\gg t_F$, we perform the perturbation in two steps. First, similarly to the previous model, we linearize the Hamiltonian close to the Fermi level. The operators $\Psi_{n\tau \lambda \eta \sigma}$ are represented in terms of left $L_{n\tau \lambda \eta \sigma}$ and right $R_{n\tau \lambda \eta \sigma}$ movers.
In the tunneling term we keep only resonant slow-varying terms,
\begin{align}
&H_1= 
t_1 \sum_{n \tau \lambda} R_{n \tau \lambda \bar 1  \lambda}
^\dagger L_{n \tau \bar \lambda  \bar 1  \lambda} + t_1 \sum_{n \tau} R^\dagger_{n \tau \bar \tau  1  \tau } L_{n \bar \tau  \tau  1  \tau}\\
&+   t_1 \sum_{n \tau}(  R^\dagger_{n \bar 1 \bar 1  1  1} L_{(n+1)  1 1  1  1}+L^\dagger_{n \bar 1 \bar 1  1  \bar 1} R_{(n+1)  1 1  1  \bar 1})+ {\rm H.c.}  \nonumber 
\end{align}
All modes appearing in $H_1$ are gapped out with a gap of size $t_1$. The remaining modes are gapped out by the periodic driving. We note that here we can use the Floquet technique for each wire independently. The corresponding Hamiltonian density is given by
\begin{align}
&H_F=t_F \sum_{n \tau \lambda \eta}   R^\dagger_{n \tau \lambda  \eta  (\eta \lambda)} L_{n \tau  \lambda \bar \eta (\eta \lambda)} +{\rm H.c.}
\end{align}
Again, the bulk modes in $H_F$  are gapped out.
However, there are two modes, one at each edge wire, that  are neither coupled  by $H_1$ nor by $H_F$ to the rest.
For example, if the system is composed out of an integer number of unit cells $N$, there are two gapless modes with opposite spins and velocities at each of the two edge wires. The most right (left) wire hosts helical edge modes $L_{(n=1)1111}$ and $R_{(n=1)111\bar 1}$ ($L_{(n=N)\bar 1 \bar 11\bar 1}$ and $R_{(n=N)\bar 1\bar 11 1}$). 
Thus, the system is in the topological regime and represents the Floquet version of a two-dimensional TI.

{\it Fractional Floquet  TI.}  The same model as in the preceding section can  also be brought into the fractional regime. To get, for example, the effective filling factor $\nu=1/3$,  we tune the chemical potential down to $E_{so}/9$ \cite{yaroslav,AQHE}. 
However, in this case, the  inter-wire tunneling is possible only if backscattering events due to electron-electron interactions are involved.  We note that Floquet representation remains unchanged also in the presence of interactions since they commute with the driving field operator. The interactions can be treated non-perturbatively by standard bosonization: $R_{n\tau \lambda \eta \sigma} = e^{i\phi_{n 1 \tau \lambda \eta \sigma}}$ and  $L_{n\tau \lambda \eta \sigma} = e^{i\phi_{n \bar 1 \tau \lambda \eta \sigma}}$, and by introducing new bosonic fields $\tilde \phi_{n r \tau \lambda \eta \sigma}=(2 \phi_{n r \tau \lambda \eta \sigma}- \phi_{n \bar r \tau \lambda \eta \sigma})/3$.  Next, we use a two-step perturbation procedure and assume that the tunneling term  describes the dominant process. This process involves the back-scattering of two electrons and is the lowest order process in $t_1$ that satisfies both spin and momentum conservation, where $g_B$ is the electron-electron back-scattering amplitude \cite{Ady_FMF,PFs_Loss,PFs_Loss_2},
\begin{align}
&H_1^{ee} = g_1 \sum_{n\tau\lambda}\cos [3(\tilde \phi_{n 1 \tau \lambda \bar 1  \lambda}-\tilde \phi_{n \bar 1\tau \bar \lambda \bar 1  \lambda})]  + g_1 \sum_{n \tau}  \\
 &  \cos [3 (\tilde \phi_{n 1 \tau \bar \tau  1  \tau} -\tilde \phi_{n \bar 1 \bar \tau \tau  1  \tau})] + g_1 \sum_{n } \big(\cos [3(\tilde \phi_{n 1 \bar 1 \bar 1  1  1}\nonumber \\
&\hspace{-10pt}-\tilde \phi_{(n+1) \bar 1  1 1  1  1})]+\cos [3(\tilde \phi_{(n+1) 1  1  1  1  \bar 1}-\tilde \phi_{n \bar 1  \bar 1 \bar 1  1  \bar 1})]\big). \nonumber 
\end{align}
Again, the driving  frequency matches the energy difference between the bands, see Fig. \ref{2D}. For weak driving it is sufficient to include the electron-electron interactions inside each band.
The term that commutes with $H_{1}^{ee}$ and satisfies the conservation laws such that it could be ordered simultaneously in the renormalization group sense  \cite{Ady_FMF,PFs_Loss,PFs_Loss_2}  leads to a gap and is given by 
\begin{align}
&H_F^{ee}=g_F \sum_{n \tau \lambda \eta} \cos [3(\tilde \phi_{n 1 \tau \lambda \eta (\eta \lambda)}-\tilde \phi_{n \bar 1 \tau \lambda  \bar \eta  (\eta  \lambda)})].
\end{align}
Here, the amplitude $g_1\propto t_1 g_B^2$ ($g_F\propto t_F g_B^2$) 
depends on the amplitude $g_B$ of back-scattering processes.
Similarly to the integer case, two helical fields at each edge (for example, $\tilde \phi_{(n=1)\bar 1 1111}$ and $\tilde \phi_{(n=1)1 111\bar 1}$ in the first unit cell)  do not enter into the coupling Hamiltonians and stay gapless. In addition, elementary excitations in these modes have fractional charge $e/3$. All this shows that we have generated a Floquet version of a {\it fractional} TI. The  same procedure can be straightforwardly generalized to other odd integer values $m$, giving rise to fractional Floquet TIs with fractional excitations $e/m$.

{\it Conclusions.} We have studied arrays of weakly coupled Rashba wires that can be brought into the topological regime by periodic driving. Our first model hosts dispersionless zero-energy modes and may be identified with Weyl semimetals. The second model is in the TI (quantum spin Hall effect) regime and hosts helical edge modes. The models can also exhibit the fractional regime where the elementary excitations have fractional charge $e/m$ with $m$ being an odd integer. The proposed effects can be measured in transport experiments or via the density of states 
\cite{Fl_Oka,Fl_Demler,Fl_Nature_Linder,Fl_PRB_Linder}. 
It is worth noting that the same models can be obtained without driving but instead by doubling the wires \cite{footnote1}.

Finally, an important issue is non-equilibrium heating and relaxation mechanisms producing mobile bulk quasi-particles. These difficulties can be overcome along the lines suggested in Refs.  \cite{Fl_Oka,Fl_Demler,Fl_Nature_Linder,Fl_PRB_Linder}, {\it e.g.}, by using adiabatic build-up of states. In addition, irrespective of the degree of disturbance the direct transitions have on the occupation of Floquet bands, the existence of edge modes can be probed directly by observing their propagation \cite{fang2012:NP} or localization \cite{kitagawa2012:NC} at the edge. 
While the goal of the present work is to provide a proof-of-principle for Floquet versions of (fractional) TIs and Weyl semimetals in the ideal case, it will be interesting to undertake  a detailed analysis of such non-ideal  effects for our particular models in future work.

\acknowledgments

We acknowledge support from the Swiss NSF and NCCR QSIT.


\newpage

\onecolumngrid

\bigskip 

\begin{center}
\large{\bf Supplemental Material to `Topological  Floquet Phases  in Driven Coupled Rashba Nanowires' \\}
\end{center}
\begin{center}
Jelena Klinovaja$^1$, Peter Stano$^2$, and Daniel Loss$^{1,2}$\\
{\it $^1$Department of Physics, University of Basel, Klingelbergstrasse 82, CH-4056 Basel, Switzerland\\
$^2$RIKEN Center for Emergent Matter Science, Wako, Saitama 351-0198, Japan}
\end{center}

\twocolumngrid

\section{ Edge Mode Wavefunctions}
In the absence of magnetic fields and if $t_F<|t_1-t_2|$, the system hosts two dispersionless zero-energy edge states at each edge. The corresponding wavefunctions at the left edge are given by
\begin{align}
\Phi_1=\begin{pmatrix}
&e^{i\phi} (e^{-x/\xi_-} -e^{-x/\xi_t} e^{2 i k_{so} x})\\
&0\\
& i e^{i\phi} (e^{-x/\xi_-} - e^{-x/\xi_t} e^{2 i k_{so}  x})\\
&0\\
&i  (e^{-x/\xi_t}e^{-2 i k_{so}  x} - e^{-x/\xi_-} )\\
&0\\
&(e^{-x/\xi_t} e^{-2 i k_{so} x}-e^{-x/\xi_-})\\
&0
\end{pmatrix}
\end{align}
and 
\begin{align}
\Phi_2=\begin{pmatrix}
&0\\
&e^{i\phi} (e^{-x/\xi_-} -e^{-x/\xi_t} e^{-2ik_{so}x})\\
&0\\
& i e^{i\phi} (e^{-x/\xi_t}e^{-2ik_{so}x} - e^{-x/\xi_-})\\
&0\\
&i  (e^{-x/\xi_-} - e^{-x/\xi_t}e^{2ik_{so}x})\\
&0\\
&(e^{-x/\xi_t} e^{2ik_{so}x}-e^{-x/\xi_-})
\end{pmatrix}.
\end{align}
in the basis ($\Psi_{111}$, $\Psi_{\bar 111}$, $\Psi_{1\bar 11}$, $\Psi_{\bar 1\bar 11}$, $\Psi_{11\bar1}$, $\Psi_{\bar11\bar1}$, $\Psi_{1\bar1\bar1}$, $\Psi_{\bar 1\bar1\bar 1}$). The localization lengths are $\xi_t=\hbar \upsilon_F/t_F$ and $\xi_- = \hbar \upsilon_F/(t_y-t_F)$.

In the presence of a magnetic field, the phase diagram is more complex. If $\Delta_Z>t_y$ and $|t_y-\Delta_Z|<t_F<(t_y+\Delta_Z)$, the zero-energy edge mode wavefunction is  given by
\begin{align}
\Phi_1 (x) = \begin{pmatrix}
e^{i\phi} (e^{-x/\xi_{1-}} -e^{-x/\xi_t} e^{2 i k_{so} x})\\
i e^{i\phi} (e^{-x/\xi_t} e^{-2 i k_{so}  x}-e^{-x/\xi_{1-}} )\\
-i e^{i\phi} (e^{-x/\xi_t} e^{2 i k_{so} x}-e^{-x/\xi_{1-}} )\\
-e^{i\phi} (e^{-x/\xi_{1-}} -e^{-x/\xi_t} e^{-2 i k_{so}  x})\\
i (e^{-x/\xi_t} e^{-2 i k_{so} x}-e^{-x/\xi_{1-}} )\\
-(e^{-x/\xi_t} e^{2 i k_{so} x}-e^{-x/\xi_{1-}} )\\
(e^{-x/\xi_t} e^{-2 i k_{so} x}-e^{-x/\xi_{1-}} )\\
-i (e^{-x/\xi_t} e^{2 i k_{so}Fx}-e^{-x/\xi_{1-}} ),
\end{pmatrix}
\end{align}
where  the localization length is defined as $\xi_{1-} = \hbar \upsilon_F/ (t_F-t_y-\Delta_Z)$.  For $t_F<|t_y-\Delta_Z|$, there is a second zero-energy edge mode given by
\begin{align}
\Phi_2 (x) = \begin{pmatrix}
i e^{i\phi} (e^{-x/\xi_{2-}} -e^{-x/\xi_t} e^{2ik_{so}x})\\
e^{i\phi} (e^{-x/\xi_t} e^{-2ik_{so}x}-e^{-x/\xi_{2-}})\\
e^{i\phi} (e^{-x/\xi_t} e^{2ik_{so}x}-e^{-x/\xi_{2-}})\\
i e^{i\phi} (e^{-x/\xi_{2-}} -e^{-x/\xi_t} e^{-2ik_{so}x})\\
(e^{-x/\xi_{2-}} -e^{-x/\xi_t} e^{-2ik_{so}x})\\
-i (e^{-x/\xi_{2-}} -e^{-x/\xi_t} e^{2ik_{so}x})\\
-i (e^{-x/\xi_{2-}} -e^{-x/\xi_t} e^{-2ik_{so}x})\\
(e^{-x/\xi_{2-}} -e^{-x/\xi_t} e^{-2ik_{so}x})
\end{pmatrix},
\end{align}
where the localization length is defined as $\xi_{2-}=\hbar \upsilon_F /(t_F-t_y+\Delta_Z)$.

\newcommand{\be}{\begin{equation}}
\newcommand{\ee}{\end{equation}}

\section{Derivation of the dynamical gap}

We consider the following microscopic single particle Hamiltonian of the electron in the nanowire 
\begin{equation}
H = \frac{{\bf p}^2}{2m_e} + V_{lat}({\bf r}) + V_w({\bf r}) + H_{so}({\bf p}, \boldsymbol{\sigma}),
\label{eq:Hc}
\end{equation}
with $m_e$ the free electron mass, ${\bf p}=-i\hbar \boldsymbol{\nabla}$ the canonical momentum operator, ${\bf r}$ the position operator, $V_{lat}({\bf r})$ the crystal lattice potential with the symmetry of the lattice, and $V_w({\bf r})$ the confinement potential defining the nanowire.
The last term is the spin-orbit interaction, 
\begin{equation}
H_{so}({\bf p}, \boldsymbol{\sigma})=\frac{g \mu_B}{4 m_e c^2} (\boldsymbol{\nabla} V_{lat}({\bf r})\times {\bf p}) \cdot \boldsymbol{\sigma},
\end{equation}
with $g=2$ the free electron g-factor, $c$ the speed of light, $\boldsymbol{\sigma}$ the vector of sigma matrices, and we neglected the electric fields generated by $V_w$ with respect to those from $V_{lat}$. We include effects of an external time dependent electric field ${\bf E}(t)$, parametrized by the vector potential through ${\bf E}(t)=-\partial_t {\bf A}(t)$. We consider a field uniform in space \cite{fnt1}, oscillating in time with frequency $\omega$, amplitude $E_0$, and linearly polarized along unit vector $\hat{\bf z}$, 
\be
{\bf E}_l(t)=E_0 \hat{\bf z} \cos\omega t.
\label{eq:El}
\ee
The most straightforward realization of such a field would be a harmonic drive of the potential difference between a back- and topgate 
enclosing the wire array.
Alternatives include, e.g., a linearly polarized electromagnetic radiation propagating along the wire array plane, or a TE mode of a waveguide \cite{karzig2015:PRX}.

Though we do not consider it in the main text, for the sake of discussion of the selection rules below, we mention also the case of a field circularly polarized along the wire axis $\hat{\bf x}$,
\be
{\bf E}_c(t)=E_0 (\hat{\bf z} \cos\omega t \pm \hat{\bf y} \sin\omega t),
\label{eq:Ec}
\ee
with $\pm$ for left and right handed polarization, respectively. Such a polarized light could be used to open (or modulate) the gap $\Delta_g$, in analogy to Ref.~\cite{gu2011:PRL}.

The electric field enters the Hamiltonian in Eq.~\eqref{eq:Hc} through the minimal coupling ${\bf p} \to {\bf p} + e {\bf A}$ with $e>0$ the elementary charge. Since the ${\bf A}^2$ term is, for a spatially uniform field, a function of time only, it can be gauged away from the total Hamiltonian \cite{fnt2},
 which then becomes a sum of the unperturbed part and the linear coupling to the field, $H=H_0+\Delta(t)$, with the latter given as
\be
\Delta(t) = e {\bf A}(t)\cdot \left( \frac{\bf p}{m_e} + \frac{g \mu_B}{4 m_e c^2} \boldsymbol{\sigma} \times \boldsymbol{\nabla} V_{lat}({\bf r}) \right).
\ee
For fields in Eqs.~\eqref{eq:El}-\eqref{eq:Ec} one can check that 
\be
{\bf A}(t)=\frac{1}{\omega}{\bf E}(t+\pi/2\omega) \equiv \frac{1}{\omega}{\bf E}(t^\prime).
\ee
Together with the operator identity obtained from Eq.~\eqref{eq:Hc} 
\be
{\bf p} = \frac{m_e}{i\hbar} [{\bf r},H-H_{so}],
\ee
we can write the coupling to the field as
\be
\Delta(t) = \frac{e}{i\hbar \omega} [{\bf E}(t^\prime)\cdot {\bf r},H],
\label{eq:delta}
\ee
which shows that in this gauge the driving field does not influence energies of the unperturbed system eigenstates.

If the nanowire dimensions are larger than the atomic scale, the eigenstates of Eq.~\eqref{eq:Hc} can be well described by the effective mass envelope function approximation. This means the effects of the crystal potential is to define bulk bands indexed by integer $b$, each parametrized with the corresponding set of Bloch wave functions $\Phi_{b,\sigma}({\bf r})$, (taken at the point ${\bf k}={\bf 0}$, with the (pseudo-)spin $\sigma_b$ according to the band degeneracy), and the band specific effective parameters (mass $m_b$, g-factor and spin-orbit interaction) which define the envelope part Hamiltonian
\begin{equation}
H_b = \frac{{\bf p}^2}{2m_b} + V_w({\bf r}) + H_{so,b}({\bf p}, \boldsymbol{\sigma}_b),
\label{eq:Hb}
\end{equation}
with $\boldsymbol{\sigma}_b$ the band (pseudo-)spin operator.
The eigenstates $\psi_{b,i}({\bf r})$ of $H_b$, which are labelled by index $i$ comprising quantum numbers corresponding to the symmetries of the wire confinement $V_w$, then complete the total wave function by the envelope part,
\be
\Psi_{b,i}({\bf r}) = \Phi_{b,\sigma}({\bf r}) \psi_{b,i}({\bf r}).
\label{eq:psi}
\ee
Here, the first function has the lattice periodicity, 
while the second is approximately constant on the atomic scale.  

The matrix element of the field between two states $\Psi_1$ and $\Psi_2$,  with energies $E_1$, and $E_2$, follows from Eq.~\eqref{eq:delta} as 
\be
\langle \Psi_1 | \Delta(t) | \Psi_2 \rangle = \frac{E_2-E_1}{i \hbar \omega}  e {\bf E}(t) \cdot {\bf d}_{12},
\label{eq:Delta}
\ee
where we defined the dipole moment 
\be
{\bf d}_{12} = \int {\rm d} V\, \Psi_1^\dagger({\bf r})  \, {\bf r} \, \Psi_2({\bf r}).
\ee
To proceed, we split the integration over the wire volume $V$ to a sum over integrals within the unit cell volumes $V_\alpha$ indexed by $\alpha$ with positions ${\bf r}_\alpha$ and use the fact that within the unit cell the envelope function is approximately constant, while the Bloch functions are periodic. We get
\be \begin{split}
{\bf d}_{12} &= \sum_\alpha \Phi_1^\dagger({\bf r}_\alpha) \Phi_2({\bf r}_\alpha) \int {\rm d} V_\alpha \, \psi_1^\dagger ({\bf r}) [  {\bf r}_\alpha + ({\bf r} - {\bf r}_\alpha) ] \psi_2({\bf r}) \\
&\approx \langle \Phi_1 |  {\bf r}| \Phi_2 \rangle\langle \psi_1 | \psi_2 \rangle + \langle \Phi_1 | \Phi_2 \rangle \langle \psi_1 |  {\bf r}| \psi_2 \rangle,
\end{split}
\label{eq:d}
\ee
where the scalar products are defined as integrals over the wire volume $V$ for the envelope functions, and over the unit cell centered at ${\bf r}=0$ for the Bloch functions, respectively \cite{fnt3}.

Equation \eqref{eq:d} elucidates how the coupling selection rules arise. 
Taking for example a cylindrical wire, the index $i$ comprises the transverse main and orbital quantum numbers $m,l$, the spin $\sigma$, and the longitudinal momentum $k_x$.
If the two states derive from the same bulk band, with $b_1=b_2$, we have $\langle \Phi_1 | \Phi_2 \rangle=1$.
If, on the other hand, the two states derive from different bulk bands, the Bloch parts are orthogonal and only the first term in Eq.~\eqref{eq:d} can contribute. The envelope functions must have the same symmetries for a non-zero overlap. (As in this scenario the masses will be in most of the cases different, the radial numbers may differ and the selection rule requires only $l_1=l_2$ and $k_{x1}=k_{x2}$.) In both cases the field polarization may impose selection rules according to angular momentum of the two states.

Let us assume for concreteness that the system we considered in the main text is realized as the conduction band being the $s$-orbital electronic band (spin $\pm 1/2$) and the valence band the $p$-orbital heavy hole band (with angular momentum $\pm 3/2$ along the wire axis $\hat{\bf x}$), and the electric field is linearly polarized along $\hat{\bf z}$. We illustrate the arising dynamical gap by considering the following block of the (infinite, see Ref. \cite{PACTs_1}) Floquet matrix, see Ref. \cite{review_1},
\be
\left( \begin{tabular}{cc|cc}
$H_c$ & 0 & 0 & $\Delta_{cv}^{(-1)}$\\
0& $H_v$  & $\Delta_{vc}^{(-1)}$ & 0\\
\hline
0 & $\Delta_{vc}^{(1)}$ & $H_c+\hbar \omega$ & 0 \\
$\Delta_{cv}^{(1)}$ & 0 &0& $H_v+\hbar \omega$\\
 \end{tabular}
\right).
\label{eq:HF}
\ee
Here the sub-blocks separated by lines correspond to photon indexes $0$ (upper) and $1$ (lower). The off-diagonal elements correspond to Fourier components defined by
\be
O^{(n)} = \frac{1}{T} \int_0^T {\rm d} t\, O(t) e^{-i n \omega t}, 
\label{eq:Fourier}
\ee
with $T=2\pi/\omega$ the driving period. We also assume that 
all quantum numbers except $\sigma$ are fixed, {\it i.e.}, $m$, $l$, $k_x$, so that each entry in Eq.~\eqref{eq:HF} is  a 2 by 2 matrix in the pseudo-spin space (that is, $\sigma\in\pm 1/2$ for the conduction band and $\sigma^\prime\in\pm 3/2$ for the valence band).
Assuming the quantum numbers respect the selection rules described above for the off-diagonal elements to be non-zero (for the specific example, the longitudinal momenta $k_{x}$ and orbital momenta $l$ along the wire axis have to be the same), the matrix elements are found from Eqs.~\eqref{eq:El}, \eqref{eq:Delta}, and \eqref{eq:Fourier} as
\be
[\Delta_{cv}^{(\pm 1)}]_{\sigma\sigma^\prime} =\pm i \frac{H_c^{\sigma\sigma}-H_v^{\sigma^\prime\sigma^\prime}}{\hbar \omega} \frac{e E_0}{2} d_{cv}^{\sigma\sigma^\prime}  \langle \psi_{c,\sigma} | \psi_{v,\sigma^\prime} \rangle, 
\label{eq:Deltacv}
\ee
with
\be
d_{cv}^{\sigma\sigma^\prime} =  \langle \Phi_{c,\sigma} | z | \Phi_{v,\sigma^\prime} \rangle.
\label{eq:dcv}
\ee
For our example, the non-zero elements are $d_{cv}^{+1/2,+3/2} = d_{cv}^{-1/2,-3/2} \equiv d_{cv}$,
the dipole moment between the conduction and valence band.

This selection rule allows us to simplify the coupling matrix elements further, assuming the envelope wave functions are factorized into the orbital and spinor parts $\psi_{i,\sigma}=\psi_i \otimes |\sigma\rangle$, to \cite{fnt4}
\be
[\Delta_{cv}^{(\pm 1)}]_{\sigma\sigma^\prime} =\pm i \frac{H_c^{\sigma\sigma}-H_v^{\sigma^\prime\sigma^\prime}}{\hbar \omega} \frac{e E_0}{2} d_{cv}  \langle \psi_c | \psi_v \rangle \langle \sigma_c | \sigma_v^\prime \rangle.
\label{eq:Deltacv2}
\ee
If, as we assumed in the main text, the spin-orbit fields in the two bands are parallel, at the degeneracy point  $H_c^{\sigma\sigma}-H_v^{\sigma^\prime\sigma^\prime} \approx \hbar \omega$
the electric field opens a gap $2t_F$ in the spectrum of the Floquet matrix with  
\be
t_F = |\Delta_{cv}^{(-1)}| \approx e E_0 d_{cv}/2.
\ee
However, we also note that in the more general case the only consequence of non-collinear spin-orbit fields is a suppression of the dynamical gap by the last term in Eq.~\eqref{eq:Deltacv2}, e.g. equal to 1/2 in case of orthogonal spin-orbit fields.

\section{Inter-wire scattering}

In addition to terms given in the main text in Eq.~(3), the electric field allows also for an inter-wire photon-assisted tunneling, described by the term (we skip the $k_y$ index which is the same on all fields)
\be
H_{F}^\prime=   \sum_{\lambda \eta \sigma} t_{F,\sigma\sigma^\prime}^\prime \Psi_{\lambda \eta \sigma}^\dagger  \Psi_{\overline{\lambda} \overline{\eta}\sigma^\prime}.
\ee
After the linearization the following spin diagonal terms survive
\begin{align}
&H_{F,\sigma\sigma}^\prime = t_{F,\sigma\sigma}^\prime \sum_{\lambda \sigma}  (R^\dagger_{\lambda (\lambda\cdot\sigma)\sigma} R_{ \overline{\lambda} (\overline{\lambda}\cdot\sigma)  \sigma  } + L^\dagger_{\lambda (\overline{\lambda}\cdot\sigma)\sigma} L_{ \overline{\lambda} (\lambda\cdot\sigma)  \sigma   }),
\end{align}
which correspond to scattering events described by the two green arrows in Fig.~2 with swapped endpoints (arrowheads). We note that this term does not result in an opening of gaps in the spectrum \cite{Bernd,Tobias} and, thus, do not change the topological criteria on the qualitative level. However, it leads to small re-normalization of the Fermi velocities and therefore of the localization lengths. 

In addition, if the spin-orbit fields are non-collinear, there is also the term
\begin{align}
&H_{F,\sigma\overline{\sigma}}^\prime = t_{F,\sigma\overline{\sigma}}^\prime \sum_{\lambda \eta \sigma}  (R^\dagger_{\lambda \eta\sigma} L_{ \overline{\lambda} \overline{\eta} \overline{\sigma}  } +  L^\dagger_{\lambda \eta\sigma} R_{ \overline{\lambda} \overline{\eta} \overline{\sigma}  }).
\end{align}
Importantly, the magnitudes $t_F^\prime$ are much smaller than the intra-wire scattering $t_F$, since they are suppressed by the same factor as is the inter-wire coupling suppressed with respect to the intra-wire coupling (the orbital part overlap in last term in Eq.~\eqref{eq:Deltacv}; this factor is small for wave function corresponding to a wire pair, as the model assumes weakly coupled wires). Therefore, the inter-wire photon assisted scattering is negligible.

\section{Tunneling current}

For the models realizing helical edge states, we need a difference in electrochemical potentials between coupled wires. If one neglects charging effects (or considers that the wires are well grounded), such difference would lead to tunneling current. Let us first estimate the magnitude of this current, considering a pair of wires detuned in electrochemical potentials by $\delta \mu$. 
To efficiently block the tunneling at the Fermi energy by the energy mismatch, the detuning has to be bigger than the tunnel coupling $t_1$ by an energy equal to $\max \{k_B T, \hbar v_F / L\}$, with $k_B$ the Boltzman constant, $T$ the temperature, and $L$ the minimum of the wire length, coherence length, and the mean free path along the wire. The detuning then results in a window of energy approximately equal to $\delta \mu$, where the wire 1 is filled and wire 2 is empty. If the spectrum is linearized, there is one to one correspondence between the states in the wire 1 and wire 2. The tunneling then transports this whole energy window in time $\hbar/t_1$ from wire 1 to wire 2, which is the tunneling current to be expected. If the temperature dominates the limit on the minimal detuning, $\delta \mu \approx k_B T$, we get the current per unit length of the wire,
\be
j \approx e \frac{t_1}{\hbar} \frac{k_B T}{\hbar v_F},
\ee
which evaluates to $I=32$ nA for a wire length of 1 $\mu$m, $T=1$ K, $v_F=10^5$ m/s, and $t_1=100$ $\mu$eV. If, on the other hand, the finite momentum resolution dominates the limit on the minimal detuning, $\delta \mu \approx \hbar v_F /L$, we get a current
\be
I = e \frac{t_1}{\hbar},
\ee
which evaluates to a similar value of $I=24$ nA for $t_1=100$ $\mu$eV. Such leakage currents were also addressed both theoretically and experimentally in biased quantum wells,
where the bias is applied to generate indirect excitons \cite{Lai,Lai2,Masha,Masha2}.

However, for nanostructures the single electron charging energies are appreciable. Approximating a nanowire pair by a classical capacitor model for parallel cylinders of radius $a$ at distance $d$, the mutual capacitance per unit length is \cite{jackson}
\be
c=\frac{\pi \epsilon_0 \epsilon_r}{{\rm arccosh} (d/2a)},
\ee
which gives the single electron charging energy
\be
E_C = \frac{1}{2}\frac{e^2}{c L}.
\ee
Choosing the relative permittivity $\epsilon_r=12.9$ of GaAs, and $d=4a$, $L=1$ $\mu$m, we get $E_C/k_B=4.2$ Kelvin. If the detuning energy is made smaller than this, which is well possible if the temperature is much smaller, we see then that the tunneling current is completely blocked by the effect of the Coulomb blockade.

\section{Fractional Weyl semimetal due to electron-electron interactions.}

In the main part, we have generalized topological insulators to the fractional regime by lowering the chemical potential correspondingly. The same can be done also for the Weyl semimetal setup considered in the main text in the presence of strong electron-electron interactions which provide the missing momenta to open up new scattering channels. For definiteness, the Fermi wavevectors are given by $\pm k_{so} (1\pm 1/3)$. In what follows we keep the notation for the fields $R_{n\lambda\eta\sigma}$ and  $L_{n\lambda\eta\sigma}$ used in the main text. The regime of the interest is defined by $t_y$ being non-zero for the momenta $k_y$ under consideration. This could be easily achieved if, for example, $t_1 \gg t_2$. If the tunneling between wires is the dominant process, which corresponds to the weak driving regime, it can be written in leading order (in the RG sense \cite{PFs_Loss_SM,PFs_Loss_2_SM,Ady_FMF_SM,yaroslav_SM,oreg_2_SM,Oreg_SM,PFs_TI_SM}) as
\begin{align}
&H_{t}^{ee}= \sum_{k_y\lambda \sigma} g_t e^{i\lambda \phi}  R^\dagger_{k_y \lambda (\lambda \sigma)  \sigma } L_{k_y  \bar \lambda (\lambda \sigma)   \sigma} \\
&\times (R^\dagger_{k_y \lambda (\lambda \sigma)  \sigma } L_{k_y \lambda (\lambda \sigma)  \sigma } ) (R^\dagger_{k_y  \bar \lambda (\lambda \sigma)   \sigma}  L_{k_y  \bar \lambda (\lambda \sigma)   \sigma} )+ {\rm H.c.}, \nonumber
\end{align}
where $g_t \propto t_y g_B^2$. The corresponding driving term that commutes with $H_{t}^{ee}$ is given by
\begin{align}
&H_F^{ee}=g_F \sum_{k_y \lambda \sigma}   R^\dagger_{k_y  \lambda  (\bar \lambda \sigma)  \sigma} L_{k_y   \lambda  ( \lambda \sigma) \sigma} \\
&\times (R^\dagger_{k_y  \lambda  (\bar \lambda \sigma)  \sigma} L_{k_y  \lambda  (\bar \lambda \sigma)  \sigma})(R^\dagger_{k_y   \lambda  ( \lambda \sigma) \sigma}L_{k_y   \lambda  ( \lambda \sigma) \sigma})+{\rm H.c.}, \nonumber
\end{align}
where $g_F \propto t_F g_B^2$. Again, we introduce first standard bosonic fields $\phi_{k_y r \lambda\eta\sigma}$ defined as $R_{k_y \lambda\eta\sigma}=e^{i \phi_{k_y 1 \lambda\eta\sigma}}$ and  $L_{k_y \lambda\eta\sigma}=e^{i \phi_{k_y \bar 1 \lambda\eta\sigma}}$. Subsequently, we switch to $\tilde \phi_{k_y r \lambda\eta\sigma} = (2 \phi_{k_y r \lambda\eta\sigma} - \phi_{k_y \bar r \lambda\eta\sigma})/3$. We note that the new fields describe quasiparticle excitations with a fractional charge $e/3$ \cite{Giom}. 
Without magnetic fields the spin is a good quantum number and each state is two-fold degenerate in the spin degree of freedom, thus we can focus only on one of the two spin components, say, $\sigma=1$. In what follows we suppress the indices $k_y$ and $\sigma$ to keep the expressions more compact.  As a result, the non-quadratic terms in the Hamiltonian become
\begin{align}
&H_{t}^{ee} = 2 g_t \sum_{\lambda}  \cos [3(\tilde \phi_{1\lambda  \lambda} -\tilde \phi_{\bar 1 \bar \lambda \lambda})+\lambda \phi],\\
&H_{F}^{ee} = 2 g_F \sum_{\lambda} \cos [3(\tilde \phi_{1\lambda \bar \lambda} - \tilde \phi_{\bar 1 \lambda  \lambda})].
\end{align}
To determine the edge modes, we need to impose vanishing boundary conditions at the system edge. To do this in the bosonization language, it is most convenient to follow the unfolding procedure \cite{Giom} by doubling the system size and halving the number of fields involved by imposing the following rule
\begin{align}
&\phi_{1 \lambda} (x) = \begin{cases}
\phi_{1 \lambda \lambda}(x), & x>0\\
\phi_{\bar 1 \lambda \lambda}(-x) , & x<0\\
\end{cases},\\
&\phi_{\bar 1 \lambda} (x) = \begin{cases}
\phi_{\bar 1 \bar \lambda  \lambda}(x), & x>0\\
\phi_{ 1 \bar \lambda  \lambda}(-x) , & x<0\\
\end{cases},
\end{align}
such that the sum of the right-moving and left-moving fields at $x=0$ is zero by construction. This yields
\begin{align}
H^{ee} = \begin{cases}
2 g_t  \sum_{\lambda} \cos [3(\tilde \phi_{1\lambda } - \tilde \phi_{\bar 1 \lambda} )+\lambda \phi], & x>0 \\
2 g_F  \sum_{\lambda}  \cos [3(\tilde \phi_{1\lambda} -\tilde \phi_{\bar 1 \bar \lambda})], & x<0
\end{cases}.
\end{align}
In a final step, we introduce the conjugated fields $\phi_\lambda$ and  $\theta_\lambda$ defined as
\begin{align}
\phi_{r\lambda} = [\phi_1 + r \theta_1+ 3\lambda(\phi_{\bar1} + r \theta_{\bar 1})]/6.
\end{align}
The non-quadratic part of the Hamiltonian assumes then the form
\begin{align}
H^{ee} = \begin{cases}
4 g_t \cos (\theta_1 ) \cos (3 \theta_{\bar 1}+\phi),  & x>0 \\
4 g_F  \cos (\theta_1) \cos (3\phi_{\bar 1}),  & x<0
\end{cases}.
\end{align}
As we can see, there is a domain wall at $x=0$ between two non-commuting fields, $[\theta_{\bar 1} (x), \phi_{\bar 1}(x')]=(i\pi/3)\ {\rm sgn} (x-x')$. Such interfaces were studied before and it was shown that the domain wall hosts zero energy states \cite{Giom,PFs_Loss_SM,PFs_Loss_2_SM,PFs_TI_SM,Ady_FMF_SM}. By analogy, the scheme could be generalized to other fractional filling factors of the form $e/m$, where $m$ is an odd integer. We emphasize that the derived energy spectrum of the edge modes is dispersionless as it does not depend on the momentum $k_y$ and thus forms a flat band. The electron-electron interaction inside this flat band, potentially, could lift this degeneracy, however, the study of this effect is beyond the scope of the present work as it requires a  numerical approach. Similarly to the non-interacting case considered in the main part, the amplitude $g_t$ depends on the momentum $k_y$, thus, it could happen that $H_{t}^{ee}$
is dominant only in a finite range of momenta. This results in two Dirac cones with edge modes existing only between them, as shown in Fig. 4b of the main text. If magnetic fields are included, also in the interacting case, the initially spin-degenerate Dirac cones split and only one single flat band emerges between them.


\begin{thebibliography}{99}


\bibitem{bib:Kane2010}
M. Z. Hasan and C. L. Kane, Rev. Mod. Phys. \textbf{82}, 3045 (2010).

\bibitem{bib:Zhang2011}
X.-L. Qi and S.-C. Zhang, Rev. Mod. Phys. \textbf{83}, 1057 (2011). 

\bibitem{bib:Hankiewicz2013}
G. Tkachov and E. M. Hankiewicz, Phys. Status Solidi B \textbf{250}, 215 (2013).

\bibitem{bib:Pankratov1985}
B. A. Volkov and O. A. Pankratov, 
Pis'ma Zh. Eksp. Teor. Fiz. \textbf{42}, 145 (1985)
[JETP Lett. \textbf{42}, 178 (1985)].

\bibitem{bib:Volkov1987}
O. A. Pankratov, S. V. Pakhomov, and B. A. Volkov, Solid State Commun. \textbf{61}, 93 (1987).

\bibitem{bib:Zhang2006}
B. A. Bernevig, T. L. Hughes, S.-C. Zhang, Science \textbf{314}, 1757 (2006). 

\bibitem{bib:Zhang2007}
M. K\"onig, S. Wiedmann, C. Brune, A. Roth, H. Buhmann,
L. W. Molenkamp, X.-L. Qi, and S.-C. Zhang, Science \textbf{318}, 766 (2007).

\bibitem{bib:Zhang2009}
A. Roth, C. Brune, H. Buhmann, L. W. Molenkamp, J. Maciejko, X.-L. Qi, and S.-C. Zhang, 
Science \textbf{325}, 294 (2009).

\bibitem{Nagaosa_2009} Y. Tanaka, T. Yokoyama, and N. Nagaosa, Phys. Rev. Lett. {\bf 103}, 107002 (2009).

\bibitem{bib:Moler2012}
K. C. Nowack, E. M. Spanton, M. Baenninger, M. K\"onig,
J. R. Kirtley, B. Kalisky, C. Ames, P. Leubner, C. Brune,
H. Buhmann, L. W. Molenkamp, D. Goldhaber-Gordon,
and K. A. Moler,  Nature Materials \textbf{12}, 787 (2013).

\bibitem{Amir_ext} S. Hart, H. Ren, T. Wagner, P. Leubner, M. Mühlbauer, C. Brüne, H. Buhmann, L. W. Molenkamp, and A. Yacoby, Nature Physics {\bf 10}, 638 (2014).

\bibitem{Leo_exp} V. S. Pribiag, A. J. A. Beukman, F. Qu, M. C. Cassidy, C. Charpentier, W. Wegscheider, and L. P. Kouwenhoven, Nature Nanotechnology {\bf 10}, 593 (2015). 





\bibitem{W_1} X. Wan, A. M. Turner, A. Vishwanath, and S. Y. Savrasov, Phys. Rev. B {\bf 83}, 205101 (2011).

\bibitem{W_2}  K.-Y. Yang, Y.-M. Lu, and Y. Ran, Phys. Rev. B {\bf 84}, 075129 (2011).

\bibitem{W_3}  A. A. Burkov and L. Balents, Phys. Rev. Lett. {\bf 107}, 127205 (2011).

\bibitem{W_4} G. Xu, H. Weng, Z. Wang, X. Dai, and Z. Fang, Phys. Rev. Lett. {\bf 107}, 186806 (2011).

\bibitem{W_Sasha}  
A.A. Zyuzin and A.A. Burkov, Phys. Rev. B {\bf86}, 115133 (2012).





\bibitem{MF_Sato} M. Sato, S. Fujimoto, Phys. Rev. B {\bf 79}, 094504 (2009).

\bibitem{MF_Sarma} R.M. Lutchyn, J.D. Sau, S. Das Sarma, Phys. Rev. Lett. {\bf 105} , 077001 (2010).

\bibitem{MF_Oreg} Y. Oreg, G. Refael, F. von Oppen, Phys. Rev. Lett. {\bf 105}, 177002 (2010).

\bibitem{alicea_majoranas_2010} J. Alicea, Phys. Rev. B {\bf 81}, 125318 (2010).

\bibitem{MF_ee_Suhas} S. Gangadharaiah, B. Braunecker, P. Simon, and D. Loss, Phys. Rev. Lett. {\bf 107}, 036801 (2011).

\bibitem{potter_majoranas_2011} A. C. Potter and P. A. Lee, Phys. Rev. B {\bf 83}, 094525 (2011).

\bibitem{Klinovaja_CNT} J. Klinovaja, S. Gangadharaiah, and D. Loss, Phys. Rev. Lett. {\bf 108}, 196804 (2012).

\bibitem{Pascal} D. Chevallier, D. Sticlet, P. Simon, and C. Bena, Phys. Rev. B {\bf 85}, 235307 (2012).

\bibitem{bilayer_MF_2012} J. Klinovaja, G. J. Ferreira, and D. Loss, Phys. Rev. B  {\bf 86}, 235416 (2012).

\bibitem{Bena_MF} D. Sticlet, C. Bena, and P. Simon, Phys. Rev. Lett. {\bf 108}, 096802 (2012).

\bibitem{Rotating_field} J. Klinovaja, P. Stano, and D. Loss, Phys. Rev. Lett. {\bf 109}, 236801 (2012).

\bibitem{Ali} S. Nadj-Perge, I. K. Drozdov, B. A. Bernevig, and A. Yazdani, Phys. Rev. B {\bf 88}, 020407(R) (2013).

\bibitem{RKKY_Basel} J. Klinovaja, P. Stano, A. Yazdani, and D. Loss, Phys. Rev. Lett. {\bf 111}, 186805 (2013).

\bibitem{RKKY_Simon} B. Braunecker and P. Simon, Phys. Rev. Lett. {\bf 111}, 147202 (2013).

\bibitem{RKKY_Franz} M. Vazifeh and M. Franz,  Phys. Rev. Lett. {\bf 111}, 206802 (2013).

\bibitem{mourik_signatures_2012}
V. Mourik, K. Zuo, S. M. Frolov, S. R. Plissard, E. P. A. M. Bakkers, and L. P. Kouwenhoven, Science, {\bf 336}, 1003 (2012). 

\bibitem{deng_observation_2012} 
M. T. Deng, C. L. Yu, G. Y. Huang, M. Larsson, P. Caroff, and H. Q. Xu, Nano Lett. {\bf 12}, 6414 (2012).

\bibitem{das_evidence_2012}
A. Das, Y. Ronen, Y. Most, Y. Oreg, M. Heiblum, and H. Shtrikman, Nat. Phys. {\bf 8}, 887 (2012). 

\bibitem{Rokhinson} L. P. Rokhinson, X. Liu, and J. K. Furdyna, Nat. Phys. {\bf 8}, 795 (2012).

\bibitem{Goldhaber} J. R. Williams, A. J. Bestwick, P. Gallagher, S. S.
Hong, Y. Cui, A. S. Bleich, J. G. Analytis, I. R. Fisher,
and D. Goldhaber-Gordon, Phys. Rev. Lett. {\bf 109}, 056803
(2012).

\bibitem{marcus_MF} H. O. H. Churchill, V. Fatemi, K. Grove-Rasmussen, M.
Deng, P. Caroff, H. Q. Xu, and C. M. Marcus, Phys. Rev.
B {\bf 87}, 241401(R) (2013).

\bibitem{Ali_exp} S. Nadj-Perge, I. K. Drozdov, J. Li, H. Chen, S. Jeon, J. Seo, A. H. MacDonald, B. A. Bernevig, and A. Yazdani, Science {\bf 346}, 602 (2014).

\bibitem{Basel_exp} R. Pawlak, M. Kisiel, J. Klinovaja, T. Meier, S. Kawai, T. Glatzel, D. Loss, and E. Meyer, arXiv:1505.06078.




%
\bibitem{barkeshli_2} M. Barkeshli, C, Jian, and X.-L. Qi, Phys. Rev. B {\bf 87}, 045130 (2012).

\bibitem{PF_Linder} N. Lindner, E. Berg, G. Refael, and A. Stern, Phys. Rev. X {\bf 2}, 041002 (2012).

\bibitem{PF_Clarke} D. Clarke, J. Alicea, and K. Shtengel, Nat. Commun. {\bf 4}, 1348 (2013).

\bibitem{PF_Cheng} M. Cheng, Phys. Rev. B {\bf 86}, 195126 (2012).

\bibitem{PF_Mong} R.  Mong, D. Clarke, J. Alicea, N.  Lindner, P. Fendley, C. Nayak, Y. Oreg, A. Stern, E. Berg, K. Shtengel, and M. P. A. Fisher, Phys. Rev. X {\bf 4}, 011036 (2014).

\bibitem{vaezi_2} A. Vaezi, Phys. Rev. X {\bf 4}, 031009 (2014).

\bibitem{PFs_Loss} J. Klinovaja and D. Loss, Phys. Rev. Lett. {\bf 112}, 246403 (2014).

\bibitem{PFs_Loss_2} J. Klinovaja and D. Loss, Phys. Rev. B {\bf 90}, 045118 (2014).

\bibitem{PFs_TI} J. Klinovaja, A. Yacoby, and D. Loss, Phys. Rev. B {\bf 90}, 155447 (2014).

\bibitem{Ady_FMF} Y. Oreg, E. Sela, and A. Stern, Phys. Rev. B {\bf 89}, 115402 (2014).





\bibitem{Fl_Oka} T. Oka and H. Aoki, Phys. Rev. B {\bf 79}, 081406 (2009).

\bibitem{Fl_Demler} T. Kitagawa, E. Berg, M. Rudner, and E. Demler, Phys. Rev. B {\bf 82}, 235114 (2010).

\bibitem{Fl_Nature_Linder} N. H. Lindner, G. Refael, and V. Galitski, Nat. Phys. {\bf 7}, 490 (2011).

\bibitem{Fl_PRB_Linder} N. H. Lindner, D. L. Bergman, G. Refael, and V. Galitski, Phys. Rev. B 87, 235131 (2013).

\bibitem{Tanaka_PRL} J. Inoue and A. Tanaka, 
Phys. Rev. Lett. {\bf 105}, 017401 (2010).

\bibitem{Fl_Rudner}M. S. Rudner, N. H. Lindner, E. Berg, and M. Levin, Phys. Rev. X {\bf 3}, 031005 (2013).

\bibitem{Fl_Liu} D. E. Liu, A. Levchenko, and H. U. Baranger, Phys. Rev. Lett. {\bf 111}, 047002 (2013).

\bibitem{Fl_Reynoso} A. A. Reynoso and D. Frustaglia, Phys. Rev. B {\bf 87}, 115420 (2013).

\bibitem{Fl_Grushin} A. G. Grushin, A. Gomez-León, and T. Neupert, Phys. Rev. Lett. {\bf 112}, 156801 (2014).

\bibitem{Sen_MF} M. Thakurathi, A. A. Patel, D. Sen, and A.  Dutta, 	Phys. Rev. B {\bf 88}, 155133 (2013).

\bibitem{Platero_2013} P. Delplace, A. Gomez-Leon, and G. Platero, Phys. Rev. B {\bf 88}, 245422  (2013).






\bibitem{Lebed} A. G. Lebed, JETP Lett. {\bf 43}, 174 (1986).


\bibitem{Yakovenko_PRB} V. M. Yakovenko, Phys. Rev. B {\bf 43}, 11353 (1991).


\bibitem{Kane_PRL} C. L. Kane, R. Mukhopadhyay, and T. C. Lubensky, Phys. Rev. Lett. {\bf 88}, 036401 (2002).


\bibitem{Stripes_PRL} J. Klinovaja and D. Loss, Phys. Rev. Lett. {\bf 111}, 196401 (2013).

\bibitem{Kane_PRB} J. C. Y. Teo and C. L. Kane, Phys. Rev. B {\bf 89}, 085101 (2014).

\bibitem{Stripes_arxiv} J. Klinovaja and D. Loss, Eur. Phys. J. B {\bf 87}, 171 (2014).

\bibitem{Stripes_nuclear} T. Meng, P. Stano, J. Klinovaja, and D. Loss, Eur. Phys. J. B {\bf 87}, 203 (2014). 

\bibitem{yaroslav} J. Klinovaja and Y. Tserkovnyak, Phys. Rev. B {\bf 90}, 115426 (2014). 

\bibitem{Neupert} T. Neupert, C. Chamon, C. Mudry, and R. Thomale,  Phys. Rev. B {\bf 90}, 205101 (2014).


\bibitem{oreg_2} I. Seroussi, E. Berg, and Y. Oreg Phys. Rev. B {\bf 89}, 104523 (2014).

\bibitem{Oreg} E. Sagi and Y. Oreg, Phys. Rev. B {\bf 90}, 201102 (2014).

\bibitem{tobias_1} T. Meng and E. Sela, Phys. Rev. B {\bf 90}, 235425 (2014).

\bibitem{AQHE} J. Klinovaja, Y. Tserkovnyak, and D. Loss, Phys. Rev. B {\bf 91}, 085426 (2015) 

\bibitem{tobias_2} T. Meng, T. Neupert, M. Greiter, and R. Thomale, Phys. Rev. B {\bf 91}, 241106(R) (2015).

\bibitem{sela} G. Gorohovsky, R. G. Pereira, and E. Sela, Phys. Rev. B {\bf 91}, 245139 (2015).

\bibitem{Gutman} R. A. Santos, C.-W. Huang, Y. Gefen, and D. B. Gutman
Phys. Rev. B {\bf 91}, 205141 (2015).



\bibitem{wang2013:S}
Y. H. Wang, H. Steinberg, P. Jarillo-Herrero, and N. Gedik, 
Science {\bf 342}, 453 (2013).
\bibitem{faisal1997:PRA}
F. H. M. Faisal and J. Z. Kami\'nski, 
Phys. Rev. A {\bf 56}, 748 (1997).

\bibitem{kitagawa2011:PRB}
T. Kitagawa, T. Oka, A. Brataas, L. Fu, and E. Demler,
Phys. Rev. B {\bf 84}, 235108 (2011).


\bibitem{review} J. H. Shirley, Phys. Rev. {\bf 138}, 979 (1965).

\bibitem{PACTs} P. Stano, J. Klinovaja, F. R. Braakman, L. M. K. Vandersypen, and D. Loss, Phys. Rev. B {\bf 92}, 075302 (2015).

\bibitem{sambe1973:PRA}
H. Sambe,
Phys. Rev. A {\bf 7}, 2203 (1973).


\bibitem{Bernd_1} B. Braunecker, G. I. Japaridze, J. Klinovaja, and D. Loss, Phys. Rev. B {\bf 82}, 045127 (2010).

\bibitem{Composite_MF} J. Klinovaja and D. Loss, Phys. Rev. B {\bf 86}, 085408 (2012).

\bibitem{SM} See Supplemental Material for details on edge mode wavefunctions, Floquet formalism, and Weyl semimetals in the fractional regime.







\bibitem{footnote1}
The  former valence band is represented by one wire and the conduction band by a subsequent one, and so on, with 
the $\mu$'s
tuned to resonance. The interwire tunneling results in  the same Hamiltonian as via Floquet driving. However, simple tunneling is more susceptible to disorder, breaking the symmetry between bands. Thus, the Floquet driving seems  experimentally more realistic.

\bibitem{fang2012:NP}
K. Fang, Z. Yu, and S. Fan,	
Nat. Photonics {\bf 6}, 782 (2012).
\bibitem{kitagawa2012:NC}
T. Kitagawa, M. A. Broome, A. Fedrizzi, M. S. Rudner, E. Berg, I. Kassal, A. Aspuru-Guzik, E. Demler, and A. G. White,
Nat. Commun. {\bf 3}, 882 (2012).

\end{thebibliography}

\begin{thebibliography}{99}

\bibitem{fnt1}
The field should be spatially uniform not to induce any additional momentum kick during photon-assisted scatterings. The requirement can be soften to the field wavelength $\lambda$ being larger than the inverse of the width of the smallest gap $t$, namely $\lambda \gtrsim \hbar v_F / t$. Taking $v_F = 10^5 $ m/s and $t$ corresponding to temperature of 1 K gives a wavelength limit $\lambda \gtrsim 760$ nm achievable even for an optical drive. 



\bibitem{karzig2015:PRX}
T. Karzig, Ch.-E. Bardyn, N. H. Lindner, and G. Refael,
Phys. Rev. X {\bf 5}, 031001 (2015).
\bibitem{gu2011:PRL}
Z. Gu, H. A. Fertig, D. P. Arovas, and A. Auerbach,
Phys. Rev. Lett. {\bf 107}, 216601 (2011).

\bibitem{fnt2}
Indeed, $A^2$ can be gauged away as a time-dependent global phase by $\Psi\to U(t) \Psi$ with $U(t)=\exp \left(-(ie^2/2\hbar m) \int^t [{\bf A}(\tau)]^2 {\rm d} \tau \right)$.

\bibitem{fnt3}
We note that the result $\langle \Psi_1 | \Delta(t) | \Psi_2 \rangle = e {\bf E}(t) \cdot {\bf d}_{12}$ follows directly from choosing the potential energy gauge for the electric field, $\Delta(t)=-e {\bf E}(t) \cdot {\bf r}$.

\bibitem{PACTs_1} P. Stano, J. Klinovaja, F. R. Braakman, L. M. K. Vandersypen, and D. Loss, Phys. Rev. B {\bf 92}, 075302 (2015).

\bibitem{review_1} J. H. Shirley, Phys. Rev. {\bf 138}, 979 (1965).

\bibitem{fnt4} 
This notation means that we treat both conduction and valence spins as a pseudo-spin 1/2, by identifying the basis states $+1/2$ with $+3/2$, and  $-1/2$ with $-3/2$ in the two bands, respectively.


\bibitem{Bernd} B. Braunecker, G. I. Japaridze, J. Klinovaja, and D. Loss, Phys. Rev. B {\bf 82}, 045127 (2010).

\bibitem{Tobias} T. Meng, J. Klinovaja, and D. Loss, Phys. Rev. B {\bf 89}, 205133 (2014).


\bibitem{Lai} L. V. Butov, C. W. Lai, A. L. Ivanov, A. C. Gossard, and  D. S. Chemla, Nature {\bf 417}, 47 (2002).

\bibitem{Lai2} C. W. Lai, J. Zoch, A. C. Gossard, and D. S. Chemla, Science  {\bf 303} 503 (2004). 

\bibitem{Masha} A. V. Nalitov, M. Vladimirova, A. V. Kavokin, L. V. Butov, and N. A. Gippius
Phys. Rev. B {\bf 89}, 155309 (2014).

\bibitem{Masha2}  P. Andreakou, S. Cronenberger, D. Scalbert, A. Nalitov, N. A. Gippius, A. V. Kavokin, M. Nawrocki, J. R. Leonard, L. V. Butov, K. L. Campman, A. C. Gossard, and M. Vladimirova, Phys. Rev. B {\bf 91}, 125437 (2015).


\bibitem{jackson}
J. D. Jackson, Classical Electrodynamics, Wiley 1975, p. 80.

\bibitem{PFs_Loss_SM} J. Klinovaja and D. Loss, Phys. Rev. Lett. {\bf 112}, 246403 (2014).

\bibitem{PFs_Loss_2_SM} J. Klinovaja and D. Loss, Phys. Rev. B {\bf 90}, 045118 (2014).

\bibitem{PFs_TI_SM} J. Klinovaja, A. Yacoby, and D. Loss, Phys. Rev. B {\bf 90}, 155447 (2014).

\bibitem{Ady_FMF_SM} Y. Oreg, E. Sela, and A. Stern, Phys. Rev. B {\bf 89}, 115402 (2014).

\bibitem{yaroslav_SM} J. Klinovaja and Y. Tserkovnyak, Phys. Rev. B {\bf 90}, 115426 (2014). 

\bibitem{oreg_2_SM} I. Seroussi, E. Berg, and Y. Oreg Phys. Rev. B {\bf 89}, 104523 (2014).

\bibitem{Oreg_SM} E. Sagi and Y. Oreg, Phys. Rev. B {\bf 90}, 201102 (2014).

\bibitem{Giom} T. Giamarchi, {\it Quantum Physics in One Dimension} (Oxford University Press, Oxford, 2004).




\end{thebibliography}
\end{document}